# Approaches to Asian Option Pricing with Discrete Dividends[*]

Jacob Lundgren[†], Yuri Shpolyanskiy[‡]

February 6, 2017

**Abstract**

The method and characteristics of several approaches to the pricing of discretely monitored arithmetic Asian options on stocks with discrete, absolute dividends are described. The contrast between method behaviors for options with an Asian tail and those with monitoring throughout their lifespan is emphasized. Rates of convergence are confirmed, but greater focus is put on actual performance in regions of accuracy which are realistic for use by practitioners. A hybrid approach combining Curran's analytical approximation with a two-dimensional finite difference method is examined with respect to the errors caused by the approximating assumptions. For Asian tails of equidistant monitoring dates, this method performs very well, but as the scenario deviates from the method's ideal conditions, the errors in the approximation grow unfeasible. For general monitoring straightforward solution of the full three-dimensional partial differential equation by finite differences is highly accurate but suffers from rapid degradation in performance as the monitoring interval increases. For options with long monitoring intervals a randomized quasi-Monte Carlo method with control variate variance reduction stands out as a powerful alternative.

# 1 Introduction

The concept of an *Asian* option payoff type consists in having the payoff function depend on some notion of the **average** of the underlying asset(s). The

---

sense in which the average is understood, then, determines the specific type of Asian option considered. While using geometric averages yields convenient mathematics, arithmetic averages tend to be favored in practice [4, 5, 21, 9, 15].

Another major distinction lies in the information upon which the average is based — the *monitoring*. Options with continuous monitoring can be studied as asymptotic cases, but options traded in actuality tend to instead have a finite set of discrete *monitoring dates* or *fixings*. The usual distinction between *call* and *put* variants applies without modification. The strike price can either be *fixed* or *floating*, where the latter type utilizes the stock price at maturity as strike.

The Asian dependence on multiple samples of the underlying is often a highly desirable characteristic in that it greatly reduces sensitivity to intentional price manipulation. Further, if an actor is already exposed to the movement of a particular asset over an interval of time, Asian options can assist in accurately hedging this risk [21].

Arithmetic Asian options are unwieldy constructs from a mathematical perspective. In the case of discrete monitoring their payoff function involves a sum of correlated log-normal variates, which is not possible to express as any currently known distribution [4]. No methods of exact analytical solution have thus yet been found, and resorting to approximations or numerical methods are the only available alternatives. Among known analytical approximations, Curran's approach [3] is believed to be quite accurate as mentioned by Haug [9]. Other approximations can also be found in Haug's book. Lord in his paper [15] derives analytical bounds for the value of arithmetic Asian options. As for the standard numerical methods for option pricing, they are affected in varying degrees by the additional complexity of the Asian payoff structure.

The exposition in this paper will revolve around the pricing of discretely monitored, fixed strike arithmetic averaging Asian options with a single underlying and general dividend structures. The placement of the monitoring dates will remain briefly undefined, but particular focus will be on options with monitoring solely as an *Asian tail* — a cluster of monitoring dates in the vicinity of option maturity. Typical examples of such tails are monitoring dates at each of the last 10 calendar or business days until (and including) the maturity date.

Allowing for general dividend structures — in particular discrete dividends paid by the underlying security — is a complication that has not been handled in the majority of previous expositions on related topics. Analytical approximations cannot deal with discrete dividends. While mentioned and theoretically handled in a few papers with numerical methods for Asian options [5, 21], their results do not discuss the impact of general dividends on convergence or performance characteristics. Večeř considers discrete dividends, but his theory is applicable only to dividends that are proportional to the underlying price, not predefined fixed dividend amounts [20]. A significant part of the motivation for this paper is an interest in the issues related to general discrete dividends and the resulting effects on numerical solution methods. The practical goal is to reveal methods that provide high computational performance both with and without discrete dividends. Another desirable feature would be for the dividend correction to have minimum possible impact on the surrounding algorithm, so as to simplify both the implementation and the testing of result consistency.

## 1.1 Numerical Complications

For lattice methods, encountering fixings breaks the essential feature of *recombination*. Usually, the computational and memory complexity of lattice methods is greatly alleviated by the Markov property of the value process of many options. Introducing Asian monitoring dates voids this property, and as a result nodes existing at a fixing will spawn individual trees which in general will not reconnect. This explosion in state space renders the method computationally unpalatable. It has been proposed [10] that discarding part of that space and recreating the lost information through interpolation when required is a viable solution.

In largely similar fashion, modifying the Black-Scholes partial differential equation (PDE) to account for the Asian payoff involves the introduction of an additional spatial dimension [5, 21]. As a result, the finite difference method also suffers a severe, if generally less explosive, increase in complexity. The standard grid structure is extended in the averaging spatial dimension, creating a three-dimensional discrete problem domain. In the case of discrete monitoring, the interaction of different levels of the averaging dimension is isolated to the monitoring instants. Between fixings the method can thus be interpreted as a set of independent finite difference methods solving a discretization of the standard two-dimensional Black-Scholes equation. The interaction is then implemented as jump conditions, again with interpolation to reconstruct unavailable data as required. A few PDE approaches [18, 20] have also been suggested to avoid the problem of increased dimensionality, and thus improve computational efficiency. Unfortunately, these techniques cannot fully cover arbitrary discrete dividends.

Among the strengths of the Monte Carlo method is its resilience to model complications [13]. Expecting it to be competitive with the more rigid numerical methods for the examined options is thus not entirely baseless. The method itself requires little adaptation to be able to valuate these products, in essence only having each trajectory keep track of its running average with respect to the fixing dates. The cost of this flexibility is a comparatively unexciting rate of convergence which is also resistant to most attempts at improvement. The implicit constant accompanying the rate of convergence is a viable target, and many variance reduction techniques have been developed for its minimization [7]. One of the rare few methods carrying the potential to affect the rate of convergence of Monte Carlo estimators consists in replacing the random sampling in trajectory generation with *low-discrepancy sequences*. In reference to the alternative labeling of such sequences as *quasi-random*, the method is often termed *quasi-Monte Carlo*. All Monte Carlo methods can be extended in a straightforward manner to account for discrete dividends, with only insignificant growth of the computation time.

## 1.2 Summary of Contributions

This paper provides a high-level description of three approaches to the pricing of arithmetic Asian options with general dividend structures. A finite difference solver using the full PDE as described by Zvan et al. [21] provides somewhat of a robust, naïve solution to examine the consistency of the other methods. A recent hybrid finite-difference approach [14] combining the standard two-dimensional Black-Scholes PDE with Curran's analytical approximation [3] is outlined and further augmented, providing a significantly faster approximative solution. Finally, the Monte Carlo method offers a structurally simple alternative which is also largely unaffected by model complications.

These approaches are studied from the perspectives of consistency, convergence and computational complexity, yielding a characterization of when each one is preferable. While the Curran-based approach can by its very nature not offer consistency, it is found to quickly render highly accurate approximations in situations fulfilling its assumptions. The explosion in computational complexity of the full PDE method as the monitoring interval grows makes it a weak candidate outside tail-type setups, despite strong convergence. The Monte Carlo approach performs consistently irrespective of option characteristics, and thus makes a strong argument for its use for options not covered by the special cases of the other approaches.

# 2 Numerical Approaches

## 2.1 Full PDE Finite Differences

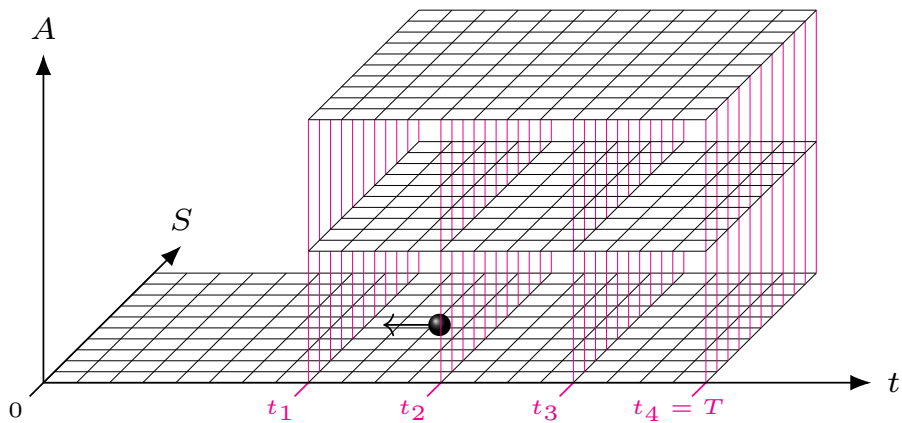

Figure 1: Structure of the full PDE FD solver

Introducing discrete Asian fixings into the Black-Scholes PDE expands the model domain into an additional spatial dimension, representing the *running* *average* process [5, 21]. Between monitoring dates, the PDE remains unchanged as no information is exchanged between levels of the running average process. Meanwhile, an argument of absence of arbitrage implies that the option value just before a fixing must equal the one just after with an appropriately corrected running average level. This creates a jump condition that is applied at the fixing dates, and the resulting process is shown in Figure 1.

A similar condition is used for discrete dividends. Since a dividend is necessarily accompanied by a drop in the value of the stock by the same amount, absence of arbitrage again implies that the value of the option as a function of the underlying stock must be shifted accordingly. Due to the structure of finite difference grids, the values required to correct for the jump conditions will in general not be available. Recreating approximations of this truncated information can be done through interpolation, which in the following will be four-point barycentric Lagrange interpolation [1].

The resulting PDE takes the form of the basic Black-Scholes PDE, with an additional parameter for the running average process:

$$\frac{\partial V}{\partial t} + \frac{1}{2}\sigma^2 S^2 \frac{\partial^2 V}{\partial S^2} + (r-\delta)S\frac{\partial V}{\partial S} - rV = 0,$$

where $V(S, A, t)$ is the option price for stock value $S$ and the current arithmetic average $A$ accumulated over passed monitoring dates $t_i \leq t$ at time $t$, $\sigma$ is the volatility; $r$ is the interest rate; and $\delta$ is the dividend (convenience) yield. Prior to the first monitoring date $t_1$, $A$ is zero. The terminal condition at maturity $T$ is given by the payoff function $\Phi(S_T, A_T)$:

$$V(S_T, A_T, T) = \Phi(S_T, A_T).$$

These functions are specific to the type of contract: call or put, i.e.

$$\Phi_{call}(S_T, A_T) = \max(A_T - K, 0),$$
$$\Phi_{put}(S_T, A_T) = \max(K - A_T, 0),$$

where $K$ is the strike price.

Letting $t_-$ denote the time an infinitesimally small time before the $N^{\text{th}}$ fixing and $t_+$ the time an infinitesimally small time after this fixing, the interaction between averaging levels is handled by the jump condition

$$V\left(S, A_{t_+}, t_+\right) = V\left(S, \frac{(N-1)A_{t_-} + S}{N}, t_-\right),$$

and letting the same notation instead denote the surroundings of a discrete dividend of size $D$, the second jump condition is

$$V\left(S_{t_+}, A, t_+\right) = V\left(S_{t_-} + D, A, t_-\right).$$

## 2.2 Hybrid Curran-Finite Differences

An approach has previously been described [14] for efficiently estimating the value of Asian options where any discrete dividend payments are performed prior to the first monitoring date. At its core is an approximation suggested by Curran [3], which introduces a conditioning on the geometric average to obtain an estimation of the value of an arithmetic Asian option on a stock not paying discrete dividends. It is possible to allow for such dividends before the first monitoring date, by setting up a standard Black-Scholes PDE with maturity at the first fixing. The approximation is then used to generate the terminal condition, and a finite difference method can be used to solve for the values at current time. The solution flow is illustrated in Figure 2.

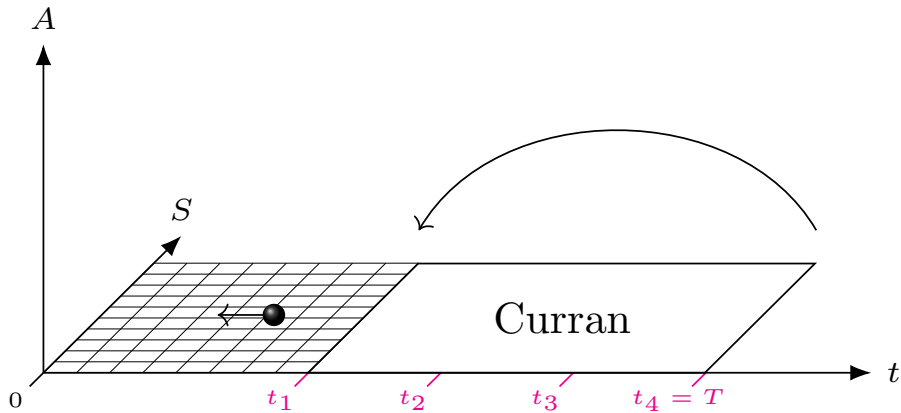

Figure 2: Structure of the Curran-based FD solver

An additional approximation allows for absolute-valued discrete dividends inside the monitoring interval. Assuming that the dividends are small in comparison to the strike, their impact can be transferred to the strike price. Let $D$ be an absolute-valued discrete dividend payment due at a time $t_D$ which is inside the monitoring interval. Denote by $t_1, t_2, \ldots, t_N$ the times of the fixings. Let further $N$ denote the total amount of monitoring dates, and $n_{\text{aff}}$ the first fixing following — and thus affected by — the dividend. With the constant interest rate $r$ and dividend yield $\delta$, the dividend effect can be compensated for by increasing the strike price by

$$K_{\text{shift}} = \frac{D}{N} \sum_{i=n_{\text{aff}}}^{N} e^{(r-\delta)(t_i - t_D)},$$

i.e. the amount by which the compounded dividend would have reduced the final arithmetic average $A_T$ via drops in spot price at future fixings. This ignores the effect of the dividend on the stock process dynamics, which introduces an error. The size of this error is one of the factors examined in the following.

## 2.3 Monte Carlo

In the case of Monte Carlo methods, the tracking of the averaging dimension is significantly less taxing. Each trajectory keeps a running average value which is updated when fixing dates are hit as shown in Figure 3, but no new dependency structures are introduced. As a result, neither the overall complexity nor the inherent parallelism of the method are affected to any considerable degree. The disappointing convergence characteristics of the approach remain an issue, however — the price of simplicity and adaptability to complications.

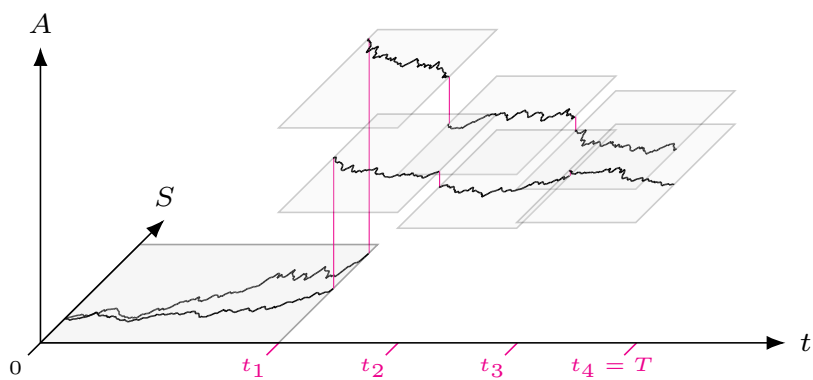

Figure 3: Structure of the Monte Carlo solver

In order to reduce the variance of the Monte Carlo estimator, a wide variety of *variance reduction techniques* have been developed [7]. While they are unable to affect the overall behavior and rate of the convergence, their use is intended to result in lower absolute variances for any given setup.

The method of *antithetic variates* consists in inverting the samples utilized to generate each trajectory and produce what is essentially a mirror image. This ensures that all scenarios of improbably high stock growth are accompanied by an equivalently low counterpart and vice versa. For options where the mapping from underlying trajectory to option value is monotone, this reduces the variance of the overall estimator.

*Control variate* methods are formalizations of the idea that knowledge of processes correlated with the option value process can be used to improve estimations of the latter. Assume that there is a process correlated with the option value process, which can be evaluated using the same underlying trajectories, and has a known expectation. This process can then be used as a control in simulations of the option value. For each trajectory, the difference between the estimate of the control process and its known expectation is used to correct the estimator of the correlated option value process.

It remains to find candidates for control processes. In the case of arithmetic Asian options, the closely related but mathematically more convenient geometric Asian options have been suggested. The value of these options is analytically known for underlyings with up to a continuous dividend yield, but since the target is to allow for general dividend structures the control is not directly usable. What is attempted is instead to simulate pairs of trajectories using the same normal samples, with one being uncorrected for discrete dividends. A geometric Asian option on the underlying without discrete dividends is then used as a control, with the assumption that the positive correlation should remain largely unaffected.

Variance reduction techniques are potent tools for reducing the implicit constant in the variance of the Monte Carlo estimator. The more problematic issue of the order 0.5 of convergence in number of trajectories, however, generally remains unaffected. A technique with the potential to improve performance in both regards, however, consists in steering the random sampling to distribute points in an artificially even manner. This can be done by the use of *low-discrepancy sequences* to feed the simulation, instead of the more common pseudo-random number generators.

Low-discrepancy (or *quasi-random*) sequences are **deterministic** constructs that derive their name from the concept of *discrepancy*, which informally can be described as a measure of how "unevenly" a set of points is distributed in a domain. Equivalently, the discrepancy can be interpreted as the amount of *clustering* the set of points exhibits. Each point added to a low-discrepancy sequence is placed to minimize the discrepancy of the resulting set of points, thus ensuring that the placement is in some sense "optimal" irrespective of the number of points used.

A wide variety of low-discrepancy sequences are available, with differing characteristics. A commonly used type is the Sobol' sequence, the generation of which is readily expressed as a highly efficient computational algorithm. The algorithm requires exogenous information both for its initialization and generation processes, and the choices affect how well the resulting sequence optimizes measures of discrepancy. This selection is anything but trivial, but authors Joe and Kuo [11, 12] have graciously made the results of their research on the matter available for public use.

To obtain samples of the normal distribution, coordinates of the points of a low-discrepancy sequence are fed into a distribution conversion algorithm such as straightforward inversion of the normal cumulative distribution function or the Box-Muller algorithm. There is debate [8, 13] as to whether the latter alternative has an undesirable impact on the distributive properties of low-discrepancy sequences. In the interest of retaining focus, the Box-Muller algorithm is used in the following for reasons of computational efficiency.

One of the advantages of standard Monte Carlo methods is the fact that the error in the estimator is asymptotically normal, allowing for estimation of the error itself by repeated invocation. With the deterministic sampling of standard quasi-Monte Carlo, however, this is not an option. In order to regain this ability, randomness can be partially reintroduced into the quasi-Monte Carlo algorithm, resulting in what is commonly termed *randomized quasi-Monte Carlo (RQMC)*. This randomization can be done in a multitude of ways, with varying computational and implementational complexity and also with different effects on the discrepancy of the resulting sequence.

One way of performing such a randomization is that of using a *random shift*. A vector of the same dimensionality as the quasi-random points is filled with uniform variates on $[0, 1)$, and added elementwise to each point in the sequence, keeping only the fractional part. It is important to note, however, that this procedure does **not** fully preserve all desirable distributional properties of the low-discrepancy sequence [16].

## 2.4 Implementation Details

The experiments detailed in the following use sequential implementations written in C++. All finite difference implementations utilize the Crank-Nicolson scheme [2] as their main method of solution. The interaction of this scheme with the class $C^0$ terminal condition and jump conditions is, however, known to result in impaired convergence rates [17, 6]. To rectify this, the suggested solution of splitting the first step into four steps of the implicit Euler scheme is used. This technique is known as Rannacher time stepping, and is applied at the start of each interval of continuity. In other words at maturity, dividend dates and — in the case of the full PDE method — at fixing dates.

The algorithm used for the Monte Carlo time stepping is direct solution of the underlying dynamics of the geometric Brownian motion driving the stock process, i.e.

$$S_{t_{i+1}} = S_{t_i} \exp\left[(r - \delta)(t_{i+1} - t_i) + \sigma\sqrt{t_{i+1} - t_i}Z\right],$$

where $t_i$ and $t_{i+1}$ are two consecutive time points and $Z \sim \mathcal{N}(0, 1)$. The ability in the standard Black-Scholes framework to perform exact stepping of this type means that taking extraneous steps shows no improvement in accuracy. The Monte Carlo methods thus step only onto dividends and fixings, with no intermediate steps.

For both the full PDE finite difference method and the Monte Carlo methods, the correction order of perfectly overlapping fixings and dividends must be defined. In a chronological sense, the used implementations apply dividend payouts before Asian monitoring, while the reverse-flowing finite difference method inverts this to obtain consistent results.

The implementation of Curran's approximation is based on the description by Haug [9], and assumes equidistant monitoring data. The impact of this approximation for non-equidistant fixing placements is one of the issues investigated in the following. The functionality for discrete dividends inside the monitoring interval also only allows compounding from the fixing following the dividend date, a fact which is assumed to generally have a significantly smaller impact.

| | | | | | | | | | | |
|---|---|---|---|---|---|---|---|---|---|---|
| *tail-eq*   | 0.9753 | 0.9781 | 0.9808 | 0.9836 | 0.9863 | 0.9890 | 0.9918 | 0.9945 | 0.9973 | 1 |
| *tail-neq*  | 0.9644 | 0.9671 | 0.9699 | 0.9726 | 0.9808 | 0.9836 | 0.9863 | 0.9890 | 0.9918 | 1 |
| *full-eq*   | 0.1    | 0.2    | 0.3    | 0.4    | 0.5    | 0.6    | 0.7    | 0.8    | 0.9    | 1 |
| *full-neq*  | 0.114  | 0.203  | 0.295  | 0.445  | 0.551  | 0.692  | 0.710  | 0.778  | 0.932  | 1 |

Table 1: Discrete monitoring times for the monitoring setups

| | | | | |
|---|---|---|---|---|
| *early* | 0.0548 | 0.3048 | 0.5548 | 0.8048 |
| *late*  | 0.2390 | 0.4890 | 0.7390 | 0.9890 |
| Size    | 2.1    | 1.9    | 2.051  | 1.949  |

Table 2: Times and sizes of discrete dividend payments for the dividend setups

# 3 Results and Discussion

## 3.1 Testing Setup

In order to understand the characteristics of the solvers, a set of varying option setups is used. The situations will in the following be labeled according to the scheme

$$<tail/full>\text{-}<none/early/late>\text{-}<eq/neq>,$$

where the fields carry the following meaning:

- $<tail/full>$ — Option fixings arranged in an Asian tail or throughout the option lifespan

- $<none/early/late>$ — Dividend placement and existence

- $<eq/neq>$ — Fixings equidistant or non-equidistant.

The setup details are shown in Tables 1 and 2, outlining respectively the monitoring times and the dividend times with associated sizes. The *tail-eq* scenarios use the last 10 calendar days for monitoring, while the *tail-neq* scenarios use the last 10 business days with two weekends contained within the monitoring interval. There is no particular rationale to the selection of the *full-neq* monitoring dates, except that the number of fixings before and after each dividend in the *late* dividend setup is consistent with the equidistant case.

The following parameters are used throughout

- Interest rate: 5% annually

- Dividend yield: 2% annually

- Volatility: 40% annually

- Maturity: 1 year

- Payoff and exercise: European Asian call

- Strike price: 100 arb. units.

The discrete dividends are fully absolute valued with the values (in arb. units) [2.1, 1.9, 2.051, 1.949].

Since the full PDE finite difference method consists only of standard numerical techniques applied to the original mathematical problem, its result on a $10^4 \times 10^4 \times 10^4$ grid is used as a reference for all error measurements. As a result, **the errors given for the full PDE method are self-referential**. Arguing for the correctness of its use as a reference thus requires the corroboration of the Monte Carlo methods converging to the same value, and the discussion must be viewed from a holistic standpoint to be considered accurate.

Stated errors for the finite difference-based solvers are given in terms of fraction of the reference value, and ignore polarity of error. The given values are the maximum relative errors in the stock interval $[0.6K, 1.4K]$, where $K$ is the strike price. For the Monte Carlo methods, only the value for stock price equal to strike price is considered, and standard deviations of the estimator are given as measures of error. **Direct comparison between the finite difference solvers and Monte Carlo solvers is thus not possible without judgment calls regarding the conversion rate between deterministic error and probabilistic deviation**.

| n | 1 | 2 | 3 | 4 |
|---|---|---|---|---|
| P | 0.682689 | 0.954500 | 0.997300 | 0.999937 |

Table 3: Probability of errors being within multiples of standard deviations

The correspondences between number $n$ of standard deviations $\sigma_{std}$ and the probability $P$ with which the error is within $(-n\sigma_{std}, n\sigma_{std})$ as given in Table 3 can be useful in deciding on desired size of standard deviation. The standard deviations are generated from sampling the Monte Carlo estimator $10^3$ times, and are shown relative to the reference value provided by the full PDE finite difference solver. The expression for the shown relative standard deviation is thus

$$\sigma_{std,rel} = \frac{1}{V_{ref}}\sigma_{std} = \frac{1}{V_{ref}}\sqrt{\frac{\sum_{i=1}^{10^3}(V_{i,MC} - V_{ref})^2}{10^3 - 1}},$$

where $V_{ref}$ is the reference value and $V_{i,MC}, i \in [1, 10^3]$ are independent samples from the Monte Carlo estimator. The subtraction of 1 in the denominator is due to Bessel's correction [19] for bias in sample variance estimators.

## 3.2 Convergence Characteristics

The convergence in all three dimensions of the full PDE finite difference solver is of second order as displayed in Figure 4 for *tail-early-eq*. The dimensions not under examination are fixed at $10^4$ nodes to isolate the error from each dimension. From the perspective of convergence, this behavior is consistent throughout all tested scenarios, with only the implicit constant changing. Using the heuristic that refinement should be in the dimension of greatest error, these plots also suggest the ratio 4 : 2 : 1 for stock:average:time nodes for this scenario. This ratio remains reasonably constant between scenarios with similar monitoring intervals (introducing a dividend shifts it slightly to 10 : 6 : 3). For options with monitoring throughout their lifespan, the ratio 10 : 3 : 3 can be found in similar fashion.

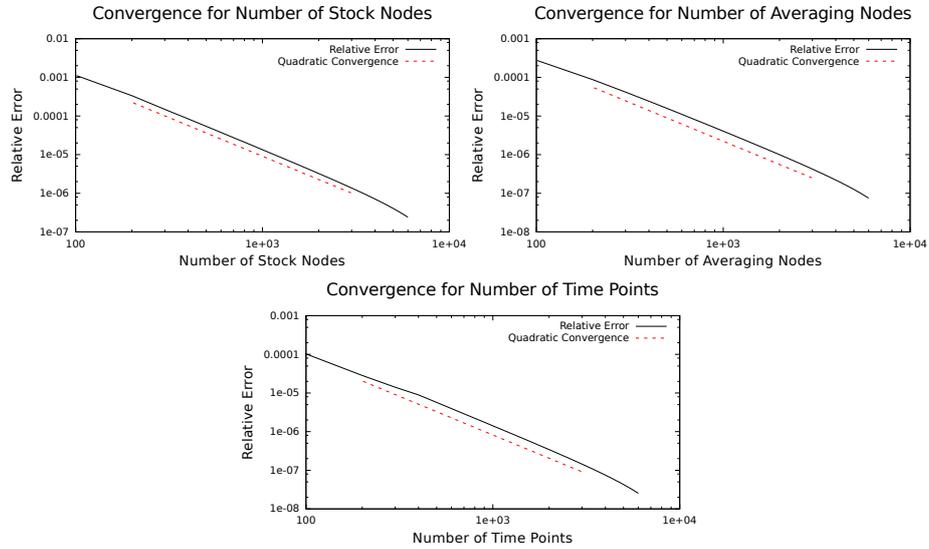

Figure 4: Orders of convergence for the full PDE FD method *(tail-early-eq)*

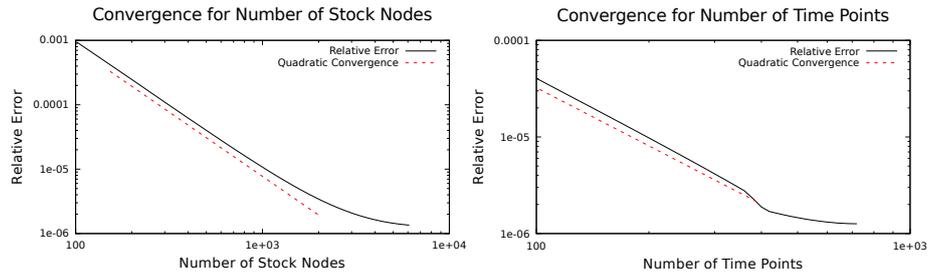

Figure 5: Orders of convergence for the Curran-based FD solver *(tail-early-eq)*

Figure 5 shows the convergence behavior of the Curran-based solution method for *tail-early-eq* — a case which interacts well with the assumptions of the approximations. The error in each dimension is isolated as above. Again, second order convergence in all dimensions is observed, until the point where the difference in converged values between the approximation and the full PDE solution becomes dominant. Due to the approximations involved, this difference is highly dependent on the considered situation, and much of the exposition in the following will be examining how the approximations fare in the different option scenarios.

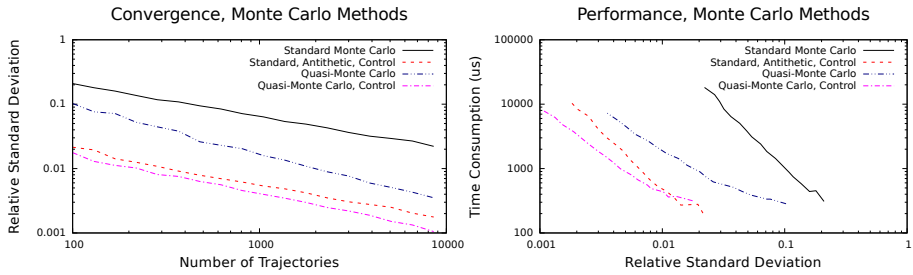

Figure 6: Convergence, performance for the Monte Carlo variants *(tail-early-eq)*

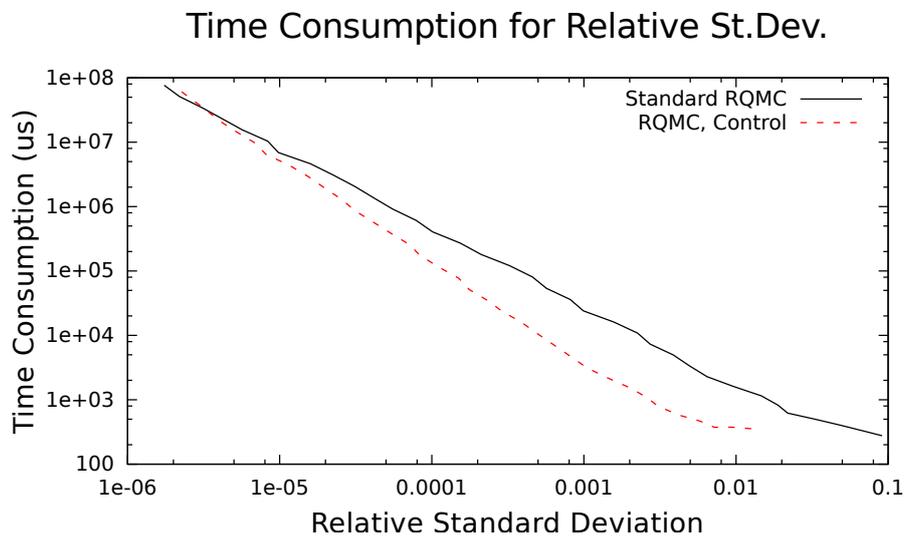

Figure 7: Behavior of standard vs controlled RQMC *(tail-late-neq)*

For the Monte Carlo variants, the first consideration is which type to use. Figure 6 shows the convergence and performance behaviors for *tail-early-eq* of the following Monte Carlo configurations

- Standard Monte Carlo without variance reduction
- Standard Monte Carlo with antithetic variates and control variates
- Randomized quasi-Monte Carlo without variance reduction
- Randomized quasi-Monte Carlo with control variates.

The observed result — the control variate RQMC method exhibiting superior behavior almost without exception — is found consistently throughout testing. Results for the Monte Carlo approach will thus be generated using this method.

It should be mentioned that the rate of convergence of the basic RQMC method is slightly greater than that of the control variate-corrected RQMC method. The error of the standard RQMC solver is $\mathcal{O}(N^{-0.79})$, while that of the controlled RQMC solver is $\mathcal{O}(N^{-0.63})$ — both of which are within the range of capacity of QMC [7]. Accordingly, despite the substantially smaller constant term for the control variate version, for sufficiently large numbers of trajectories it is overtaken performance-wise by the standard RQMC solver as shown in Figure 7. This figure also helps substantiate the claims to correctness and high accuracy of both the full PDE finite difference method used as a reference and the Monte Carlo methods themselves.

## 3.3 Error from Approximations

| Scenario | *tail-none-eq* | *tail-none-neq* | *full-none-eq* | *full-none-neq* |
|---|---|---|---|---|
| Rel. Error | 5.1e-6 | 2.9e-3 | 1.1e-2 | 1.1e-1 |
| Time ($\mu$s) | 5.8e-1 | 5.8e-1 | 6.7e-1 | 6.7e-1 |

Table 4: Curran approximation errors and time consumption

| Scenario | *tail-early-eq* | *tail-early-neq* | *tail-late-eq* | *tail-late-neq* | *full-late-eq* | *full-late-neq* |
|---|---|---|---|---|---|---|
| Rel. Error | 1.7e-6 | 3.3e-3 | 6.1e-5 | 3.2e-3 | 9.5e-2 | 2.6e-2 |
| Time ($\mu$s) | 9.3e3 | 4.6e2 | 1.3e4 | 4.1e2 | 8.5e1 | 2.4e2 |

Table 5: Curran-based FD solver errors and time consumption

Understanding the impact of the approximations involved in the Curran-based solver is imperative in determining its usefulness. Two categories of scenarios are treated separately — those with discrete dividends and those without. For the scenarios without discrete dividends, the Curran approximation can be used without modification, yielding the errors shown in Table 4. The scenarios containing discrete dividends, however, require the finite difference extensions previously outlined. Error bounds for these cases, as well as the time required to reach them, are shown in Table 5. As shown by the case *tail-late-eq*, the described correction for dividends inside the monitoring interval renders accurate results when the monitoring period is short, and the effects of the dividend on the stock dynamics thus are limited.

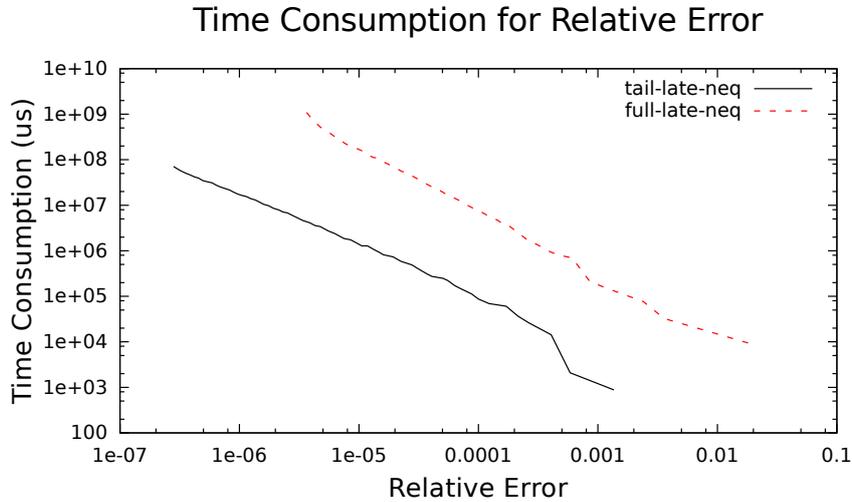

Figure 8: Effects of monitoring interval on full PDE solver performance

While the question of what is acceptable accuracy is highly dependent on application, it seems difficult to ignore the percent-magnitude errors incurred by Curran's approximation for Asian options with fixings distributed throughout their lifespan. Inversely, only those with highly specialized requirements (and great faith in the model) should be unsatisfied by the numbers posted for the scenarios with equidistant fixings in an Asian tail. The areas where judgment calls will be more diverse are the borderlands consisting of non-equidistant fixings in an Asian tail formation, as accuracy takes a very real hit from such fixing layouts.

## 3.4 Performance

From a performance and accuracy standpoint, the full PDE finite difference solver is generally unconcerned about the specifics of dividend and fixing placement and sizes, but the particular issue of monitoring interval size has an immense impact on the execution time. Figure 8 displays the difference in execution time required to reach different degrees of accuracy for an Asian tail case and a full Asian case. Due to the fact that the averaging dimension collapses to a single level $A = 0$ outside monitoring intervals, the full PDE method is significantly more palatable in scenarios with short such intervals.

There is an important caveat regarding the data displayed in Figure 8 which is related to the error estimation. The error shown is the worst case scenario, i.e. the sum of the largest estimated errors in each dimension for the setup. The choice of run parameters for the next datapoint is done by refining the dimension for which the estimated error is currently the largest.

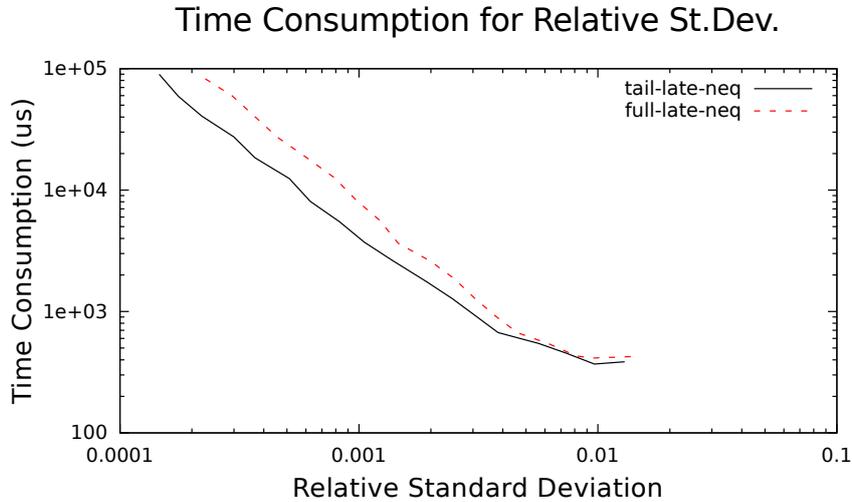

Figure 9: Effects of monitoring interval on MC solver performance

In contrast to the full PDE FD solver, the Monte Carlo solvers are affected comparatively little by changing the monitoring interval, as Figure 9 shows. Figures 8 and 9 are highly useful in establishing regions of usefulness for the different methods. Due to the different measures of error, rigorous comparative statements are again impossible. Irrespective of the conversion mapping between the errors, however, it would not seem unreasonable to say that the Monte Carlo methods are at least competitive — if not superior — for options with monitoring dates throughout their lifespan. It is also clear that the Monte Carlo methods have a significant amount of ground to make up through their inherent parallelism to be considered a contender for Asian tail scenarios.

## 4 Conclusion

Different approaches to the pricing of arithmetic Asian options with discrete, absolute dividends were tested from the perspectives of convergence and performance. It is clear that the best performance is obtained only with knowledge of specifics regarding the target option.

An option with an equidistant Asian tail benefits very little from the more general approaches using Monte Carlo or solution of the full PDE. Irrespective of dividend placement this case generally yields favorable results for the approach combining Curran's approximation with finite difference extensions. Introducing irregularity into the fixing placement is a more problematic issue, and the improved accuracy of a finite difference method solving the full PDE becomes potentially desirable without a catastrophic loss of performance.

Options with monitoring performed over a longer time period are more problematic still, with the errors in Curran's approximation becoming highly signif-icant. Meanwhile, the cost of solving the full PDE grows dramatically, and makes obtaining any reasonable degree of accuracy painful. While it could be argued that the probabilistic nature of their convergence is undesirable, Monte Carlo methods step up in these cases with their usual irreverence of increases in dimensionality. Across a wide range of desired accuracies, then, the approach of randomness (or quasi-randomness, as it were) makes a strong argument for its existence.

Of further interest is the fact that the discussion thus far has neglected another of the main strengths of Monte Carlo methods: its inherent parallelism. While it is unlikely to carry the approach to superiority in the case of Asian tails, it strengthens the argument that long monitoring intervals are the domain of Monte Carlo.

# Acknowledgements

Jacob Lundgren would like to express gratitude to Uppsala University's Professor Maciej Klimek for his guidance in the thesis project upon which parts of this work is based. His suggestions on direction and focus were highly valuable and were what allowed some of the more interesting results to be obtained.

Yuri Shpolyanskiy would like to thank Jonas Hansbo, Itiviti's CSO and colleagues at Itiviti's Engineering: Dmitriy Ivanov, Dmitriy Kislin and Alexander Toropov for stimulating discussions.

# Declaration of Interest

This work has been done largely as part of normal operations at Itiviti. The authors would argue that this affiliation is of no detriment to the objectivity of the exposition, but the fact should be mentioned. The authors alone are responsible for the content and writing of the paper.